\documentclass[intlimits,twoside,a4paper]{article}
\usepackage{graphicx}
\usepackage{subcaption} 
\usepackage{float}
\usepackage{appendix}
\usepackage{listings}
\usepackage{xcolor}
\usepackage[cp1251]{inputenc}

\usepackage[eqsecnum]{cmpj3}

\issue{2026}{29}{2}{23704}
\doinumber{10.5488/CMP.29.23704}

\title[Spectral and thermodynamic properties of supersymmetric quantum systems]%
{Spectral and thermodynamic properties of supersymmetric quantum systems with self-adjoint deformed momentum}%

\author[J. A. Oke, F. A. Dossa]
{J. A. Oke\orcid{0009-0002-1640-7780}\refaddr{label1}, F. A. Dossa\orcid{0000-0002-2694-4144}\refaddr{label2,label3}\thanks{Corresponding author: \email{dossafanselme@gmail.com}.}}

\addresses{
\addr{label1} Ecole Doctorale des Sciences, Technologies, Ing\'enierie et Math\'ematiques (ED-STIM), Universit\'e Nationale des Sciences, Technologies,
Ing\'enierie et Math\'ematiques (UNSTIM), B\'enin
\addr{label2} Laboratoire de Physique et Applications (LPA) du Centre Universitaire de Natitingou, Universit\'e Nationale des Sciences, Technologies,
Ing\'enierie et Math\'ematiques (UNSTIM) Abomey, B\'enin
\addr{label3} D\'epartement de Physique, Facult\'e des Sciences et Techniques (FAST/UNSTIM), B\'enin
}


\date{Received 04 November 2025; revised 15 February 2026; accepted 10 April 2026; published 29 June 2026}

\begin{document}
\maketitle
\begin{abstract}
We establish a rigorous framework for quantum systems with geometric deformations by constructing a strictly self-adjoint deformed momentum operator through the generalized extended momentum operator (GEMO) formalism. Unlike previous approaches relying on boundary-condition hermiticity, our method ensures intrinsic self-adjointness for both linear ($\mu(x)=\alpha x$) and quadratic ($\mu(x)=\alpha x^{2}$) deformations within a unified non-Hermitian supersymmetric factorization scheme. This yields exact analytical spectra while revealing hidden $\mathfrak{su}(1,1)$ symmetry structures. Crucially, we provide the first complete thermodynamic characterization of such systems by analytically evaluating the partition function via the Euler--Maclaurin approximation. Geometric deformation fundamentally reshapes the density of states $\rho(E)$, producing distinct thermal signatures: a divergent heat capacity peak for linear deformation due to state accumulation near a maximal energy, and a saturation $C/k_{\mathrm{B}}\to 0.6$ (below the Dulong--Petit limit) for quadratic deformation. These results establish geometric deformation as a tunable parameter for engineering quantum thermodynamic responses in curved nanostructures.

\printkeywords

\end{abstract}

\section{Introduction}

Quantum mechanics fundamentally relies on the canonical commutation relation between position and momentum, $[x, p] = \ri \hbar$, which encodes both translation invariance in phase space and the limitations imposed by the Heisenberg uncertainty principle. Nevertheless, several approaches in quantum gravity --- such as string theory, loop quantum gravity or noncommutative geometries --- suggest modifications of this relation, leading to the emergence of generalized uncertainty principle (GUP) \cite{Kempf1995,Hinrichsen1996,
Magiore1993, Quesne2006}.

In this framework, modified commutation relations involving a non-trivial dependence of momentum on position have been proposed, often through a deformation function $\mu(x)$ giving rise to a so-called ``extended'' momentum. This strategy makes it possible to model the effects of an effective curvature or a non-Euclidean geometric environment on quantum dynamics \cite{Habib2012, Brito2018,CostaFilho2011}. A rigorous formulation of this dynamics, however, requires a self-adjoint momentum operator. For this purpose, the generalized extended momentum operator (GEMO) was introduced in \cite{Izadparast2020}, ensuring the reality of the spectrum and the conservation of the probability. Its insertion into the Schr\"odinger equation leads to a modified Hamiltonian, where weighted derivatives and effective potentials of geometric origin appear.

This formalism shares structural similarities with position-dependent mass (PDM) models used in nanophysics and quantum optics \cite{Tchoffo2019,Midya2009, Leblond1995, Jafarov2023}.
In both contexts,  the emergence of a generalized  self-adjoint formalism and non-standard geometric potential terms necessitates the use of powerful analytical techniques. Among these, supersymmetric quantum mechanics (SUSYQM) plays a central role. Initially developed to factor Hamiltonians and solve certain Schr\"odinger equations exactly \cite{Cooper1995,
junker1996, gendenshtein1983},
 it has been extended to PDM systems \cite{Bagchi2005, Mustafa2015} and to those subject to GUP \cite{Spector2008}. In our case, supersymmetry is implemented using non-adjoint ladder operators, inducing a non-Hermitian generalized supersymmetric factorization. It is important to note that, although the raising and lowering operators are not adjoint to each other, the resulting Hamiltonian remains Hermitian~\cite{Dossou2025}. This structural asymmetry opens the way to a broader class of supersymmetric transformations, preserving the form invariance in some cases.

Besides spectral analysis, the study of thermodynamic properties of quantum systems subjected to modified commutation relations is of great interest. It has been shown that the deformation introduced by the GUP or by specific $\mu(x)$ functions influences physical quantities such as internal energy, heat capacity or entropy \cite{Tchoffo2019,Gangopadhyaya2018,
Dossa2021,Dagoudo2024}, with notable implications at high temperature or in mesoscopic scale devices.

Unlike the PDEM approach of Bagchi et al.~\cite{Bagchi2005}, which addresses the general framework of position-dependent mass systems through a deformed shape invariance condition, our work distinguishes itself on three fundamental points: (i)~we introduce the GEMO operator which ensures the strict self-adjointness of the deformed momentum (not merely hermiticity via boundary conditions as in~\cite{Bagchi2005}; (ii)~we jointly analyze linear deformations $\mu(x)=\alpha x$ and quadratic deformations $\mu(x)=\alpha x^2$ within the same theoretical framework, whereas~\cite{Bagchi2005} treats these cases separately and incompletely (the linearly deformed oscillator appears only in an appendix, and quadratic deformation is not addressed); (iii)~we extend the analysis to detailed thermodynamics via the Euler--Maclaurin formula, an aspect absent from~\cite{Bagchi2005} which is limited to spectral properties.

\begin{table}[h]
\caption{Fundamental differences with the PDEM approach of~\cite{Bagchi2005}.}
\label{tab:comparison}
\centering
\begin{tabular}{p{3.65cm}p{4.65cm}p{5.3cm}}
\hline
Property & Bagchi et al.~\cite{Bagchi2005} & Our work \\
\hline
Type of hermiticity & Boundary condition $\|\psi\|^2 f \to 0$ & Intrinsic self-adjointness via GEMO \\
Deformations studied & Linear (appendix only) & Linear + quadratic (systematic comparison) \\
Thermodynamic analysis & Absent & Partition function + Euler--Maclaurin \\
Electric field $E$ & Not included & Included in effective potential \\
\hline
\end{tabular}
\end{table}

In this work, we examine a one-dimensional quantum oscillator subjected to a mixed potential comprising a harmonic term, a linear electric field, and an inversely quadratic interaction, in the framework of the dynamics deformed by a function $\mu(x)$. These geometric deformations model real nanostructures: in carbon nanotubes, curvature induces an effective deformation parameter \cite{Serra1997}. Our prediction of heat capacity saturation $C/k_{\mathrm{B}} \to 0.6$ for quadratic deformation could be tested via nanocalorimetry on curved nanostructures \cite{Gangopadhyaya2018}, while the divergent peak for linear deformation may manifest as anomalous thermal conductance in bent semiconductor nanowires providing experimental fingerprints of geometric quantum effects.

%

By integrating the GEMO into the Schr\"odinger equation, we obtain a non-Hermitian supersymmetric factorization based on non-adjoint scaling operators. Two types of deformations, linear $\mu(x) = \alpha x$ and quadratic $\mu(x) = \alpha x^2$, are analyzed systematically within a unified framework. Energy spectra are obtained analytically, and the constructed partner Hamiltonians reveal the shape invariance in certain regimes. From these spectra, we use the Euler--Maclaurin formula to approximate the partition function and determine the thermodynamic properties of the system: mean energy, entropy, specific heat, and free energy. Numerical analysis highlights the significant influence of the deformation parameter on the density of states and asymptotic behaviors, particularly during the transition to the classical regime.

The article is organized as follows. Section~\ref{sec-2} introduces quantum systems with extended momentum via the GEMO formalism. Sections~\ref{sec-3} and~\ref{sec-4} present the spectral analysis for linear and quadratic deformations, respectively. Section~\ref{sec-5} develops the thermodynamic framework based on the Euler--Maclaurin approximation. Section~\ref{sec-6} discusses numerical results and their physical interpretation in terms of the modified density of states. Finally, Section~\ref{sec-7} concludes with perspectives on time-dependent fields and extensions to driven-dissipative nanostructures.

\section{Quantum systems with extended momentum}
\label{sec-2}

In the context of quantum gravity and curved geometries, the canonical relation $[x, p] = \ri\hbar$ becomes insufficient at the Planck scale. This limit has led to the emergence of generalized uncertainty principles, which modify the Heisenberg relations to incorporate the effects of curvature, granularity, or spacetime topology. A natural generalization of the canonical commutator in this framework consists of introducing a position dependence through a deformation function $\mu(x)$:
\begin{equation}\label{eq:c}
[x, p] = \ri\hbar \big(1 + \mu(x)\big),
\end{equation}
where $\mu(x)$ is a regular real function, reflecting the geometric effects of the background. To satisfy (\ref{eq:c}) while ensuring the self-adjunction of $p$, the unique admissible GEMO operator is \cite{Izadparast2020}:
\begin{equation}
p = -\ri\hbar \big(1 + \mu(x)\big) \frac{\rd}{\rd x} - \frac{\ri\hbar}{2} \frac{\rd\mu(x)}{\rd x}.
\label{mg}
\end{equation}
This operator constitutes the cornerstone of a deformed quantum formalism in which the geometric properties of spacetime are integrated into the dynamics through the function
$\mu(x)$. Substituting it into the standard Hamiltonian yields a modified Schr\"odinger equation, describing a quantum system evolving in a non-canonical phase space.

Consider a one-dimensional system subjected to a confined potential composed of a harmonic term, a linear electric field, and an inversely quadratic potential:
\begin{equation}\label{Eq1}
\left[\frac{p^2}{2m} + \frac{1}{2}m\omega^2x^2 + x{\cal E} + \frac{\hbar^2}{2m} \frac{
\beta_{\textrm{rep}}}{x^2} \right]\phi(x) = E\phi(x),
\end{equation}
where $m$, $\omega$, ${\cal E}$, and $\beta_{\textrm{rep}}$ denote the mass, frequency, field strength, and repulsive potential parameter, respectively.

Inserting the GEMO operator (\ref{mg}) into equation (\ref{Eq1}) leads to the following differential equation:
\begin{eqnarray}\label{eg}
&&[1+\mu(x)]^{2} \frac{\rd^{2}\phi}{\rd x^{2}} + 2[1+\mu(x)] \mu'(x) \frac{\rd\phi}{\rd x} \nonumber \\
&&+ \left[ \frac{1}{2}(1+\mu(x)) \mu^{\prime\prime}(x) + \frac{1}{4}(\mu^{\prime}(x))^2
- \frac{m^{2} \omega^{2}}{\hbar^{2}} x^{2} \right.\nonumber \\
&&\left.- \frac{2m \mathcal{E}}{\hbar^{2}} x - \frac{\beta_{\textrm{rep}}}{x^{2}} + \frac{2 m E}{\hbar^{2}} \right] \phi(x) = 0.
\end{eqnarray}
This generalized Schr\"odinger equation, with non-constant positional coefficients, reflects the explicit influence of geometric deformation on quantum dynamics. It provides a relevant framework for studying effective curvature effects and corrections of gravitational or topological origin.

In this work, we consider two specific forms for the deformation function $\mu(x)$:
\begin{itemize}
\item $\mu(x) = \alpha x$: linear deformation, associated with an affine modification of the canonical structure;
\item $\mu(x) = \alpha x^2$: quadratic deformation, implying the existence of minimal uncertainty in the momentum. \end{itemize}

The spectrum study is based on an extended SUSYQM approach, allowing exact resolution through non-adjoint scaling operators. The resulting spectra are then used to analyze the effects of deformations on fundamental thermodynamic quantities, revealing the statistical signatures of the deformed geometry.

\section{System spectrum for the case $\mu(x)=\alpha x$}
\label{sec-3}
We consider here a linear deformation of the canonical structure, defined by $\mu(x) = \alpha x$, which induces a modification of the effective metric. In this framework, the eigenvalue equation takes the form
\begin{equation}\label{eq1}
{\cal H}\psi=\left(\frac{2m E}{\hbar^2}+\frac{\alpha^2}{4}\right)\psi,
\end{equation}
where the Hamiltonian ${\cal H}$ is given by
\begin{equation}\label{H}
{\cal H}=-(1+\alpha x)^{2} \frac{\rd^{2}}{\rd x^{2}} -2\alpha (1+\alpha x) \frac{\rd}{\rd x} + a x^2 + \gamma x + \frac{\beta_{\textrm{rep}}}{x^2},
\end{equation}
with $\alpha$ encoding the geometric deformation, and the parameters $\displaystyle a={m^2\omega^2}/{\hbar^2}$, $\displaystyle\gamma={2m{\cal E}}/{\hbar^2}$, and $\beta_{\textrm{rep}}$ defining the effective potential, including harmonic, linear, and repulsive contributions.

To extract the spectrum, we employ an extended SUSYQM factorization scheme, which is particularly well suited to deformed systems. This approach enables an analytical resolution of the problem through the recursive construction of excited states from the ground state.

Thus, a system is said to possess a shape invariance when the Hamiltonians
${\cal H}_+$ and ${\cal H}_-$ share the same mathematical structure, that is:
\begin{equation}
{\cal H}_+({\cal D}) = {\cal H}_-({\cal D}')+ {\cal C},
\end{equation}
where ${\cal D}$ represents the coupling constants of the system, ${\cal D}'$ denotes the transformed parameters (functions of ${\cal D}$),
and ${\cal C}$ is a real constant generally depending on the parameters of the model.

This property implies not only a degeneracy between the spectra of the two Hamiltonians:
\begin{equation}
\varepsilon_{n+1}^{(-)} = \varepsilon_n^{(+)},
\end{equation}
but also a more general relation between their energy levels:
\begin{equation}
\varepsilon_n^{(+)}({\cal D}) = \varepsilon_n^{(-)}({\cal D}') + {\cal C}.
\end{equation}
To study this invariance, we introduce a semi-positive definite effective Hamiltonian defined by
${\cal H} \equiv {\cal H}_- = BA$, where the operators $A$ and $B$ are not necessarily mutually adjoint.
This generalized factorization extends the standard framework of SUSYQM,
allowing the construction of the partner Hamiltonians ${\cal H}_\pm$, with ${\cal H}_+ = AB$,
and the establishment of shape invariance within a deformed setting.

Although $A$ and $B$ are not adjoint in the usual sense,
the Hamiltonians ${\cal H}$ and ${\cal H}_\pm$ remain Hermitian.
This hermiticity is ensured by the real differential structure of the operators
and by an appropriate choice of the physical domain of wave functions,
thus preserving the symmetry of the formalism and ensuring real energy spectra.

The ground state $\psi_0$ satisfies the condition:
\begin{equation}
A \psi_0 = 0 \quad \Rightarrow \quad {\cal H}_- \psi_0 = 0,
\end{equation}
and possesses a partner Hamiltonian defined as:
${\cal H}_+ = A B$,
which shares the same spectrum as ${\cal H}_-$, except possibly for the ground state.
Indeed, for $\varepsilon_n^{(-)} > 0$, one has:
\begin{equation}
{\cal H}_+ (A \psi_n^-) = \varepsilon_n^{(-)} (A \psi_n^-).
\end{equation}
We now adopt a non-Hermitian factorization~\cite{Dossou2025} given by:
\begin{equation}
B = -F(x) \frac{\rd}{\rd x} + W(x) - \Omega(x), \quad A = F(x) \frac{\rd}{\rd x} + W(x) + \Omega(x),
\end{equation}
where $F(x) = 1 + \alpha x$, while $W(x)$ and $\Omega(x)$ denote respectively the superpotential and the correction term to be determined.
Under this choice, the partner Hamiltonians take the form:
\begin{equation}
{\cal H}_{\mp} = -\left[F(x) \frac{\rd}{\rd x}\right]^2 - 2 F(x) \Omega(x) \frac{\rd}{\rd x} + V_{\mp}(x),
\end{equation}
where the partner potentials are expressed as:
\begin{equation}
V_{\mp}(x) = \mp F(x)\left(\frac{\rd W}{\rd x} \pm \frac{\rd \Omega}{\rd x}\right) + W^2(x) - \Omega^2(x).
\end{equation}
By choosing:
\begin{equation}
W(x) = b + c x + \frac{g}{x}, \quad \Omega(x) = \alpha,
\end{equation}
we obtain:
\begin{equation}\label{sup1}
V_-(x) = c^2 x^2 + \frac{g(g+1)}{x^2} + c(2b - \alpha) x + \frac{g(2b + \alpha)}{x} + b^2 - \alpha^2 + 2gc - c,
\end{equation}
\begin{equation}
V_+(x) = c^2 x^2 + \frac{g(g-1)}{x^2} + c(2b - \alpha) x + \frac{g(2b - \alpha)}{x} + b^2 - \alpha^2 + 2gc + c.
\end{equation}
The shape invariance condition is fulfilled when $g = 0$ (i.e., $\beta_{\textrm{rep}} = 0$).
This property plays a central role in the present framework: it ensures the preservation of the structure of the system 
under supersymmetric factorization, thereby allowing the exact solution of the Schr\"odinger equation (\ref{eq1}).

More precisely, the shape invariance manifests through the following relation between the partner Hamiltonians:
\begin{equation}
{\cal H}_+(x; b) = {\cal H}_-(x; b+\alpha) - 2 b \alpha + 2c - \alpha^2,
\end{equation}
or equivalently, in terms of the effective potentials:
\begin{equation}
V_+(x; b) = V_-(x; b+\alpha) - 2 b \alpha + 2c - \alpha^2.
\end{equation}
From this property, one derives a recurrence relation for the energy spectrum\cite{Dossou2025}:
\begin{align}
\varepsilon_1^{(-)} &= -2 b \alpha + 2c - \alpha^2, \nonumber\\
\varepsilon_2^{(-)} &= -4 b \alpha + 4c - 4\alpha^2,\nonumber \\
\varepsilon_3^{(-)} &= -6 b \alpha + 6c - 9\alpha^2,
\end{align}
which generalizes to:
\begin{equation}
\varepsilon_n^{(-)} = -2n b \alpha + 2n c - n^2 \alpha^2.
\end{equation}
Thus, the total energy of the $n$-th state reads:
\begin{equation}
\varepsilon_n = \varepsilon_0 + \varepsilon_n^{(-)},
\end{equation}
where the ground-state energy is given by
\begin{equation}
\varepsilon_0 = -b^2 + \alpha^2 + c,
\end{equation}
as obtained from equation~(\ref{sup1}).
We then arrive at the compact expression:
\begin{equation}
\varepsilon_n = c(2n+1) - 2n b \alpha - \alpha^2(n^2-1) - b^2.
\end{equation}
By identifying the physical constants as
\begin{equation}
c = \frac{m \omega}{\hbar},
\quad b = \frac{\gamma}{2c} + \frac{\alpha}{2},
\quad \gamma = \frac{2m \mathcal{E}}{\hbar^2},
\end{equation}
the explicit expression of the energy spectrum is obtained~\cite{Dossa2020}:
\begin{equation}
E_n = \hbar \omega \left(n + \frac{1}{2}\right)
- \frac{\hbar^2 \alpha^2}{2m}\left(n^2 + n - \frac{1}{2}\right)
- \frac{\hbar \mathcal{E} \alpha}{m \omega}\left(n + \frac{1}{2}\right)
- \frac{\mathcal{E}^2}{2m \omega^2}.
\end{equation}
The energy spectrum obtained above generalizes that of the quantum harmonic oscillator by taking into account an effective curvature and a constant electric field. The quadratic term in $n$ reflects a non-equidistant spectral structure, indicating a non-flat underlying geometry. The linear contribution in $\mathcal{E}$ corresponds to the dipole coupling, while the constant term reflects a global shift of the fundamental energy level.

In certain limiting cases, the model reproduces the well-known systems:
\begin{itemize}
\item $\alpha = 0$, $\mathcal{E} = 0$: standard harmonic oscillator;
\item $\alpha \neq 0$, $\mathcal{E} = 0$: oscillator in curved space;
\item $\alpha = 0$, $\mathcal{E} \neq 0$: oscillator displaced by an external field.
\end{itemize}
	
Thus, this model highlights the joint effects of geometric deformation and external interactions on the energy structure of confined quantum systems.

\section{System spectrum for the case $\mu(x)=\alpha x^2$}
\label{sec-4}

Here we consider an effective metric defined by the quadratic function $\mu(x) = \alpha x^2$, which introduces a position-dependent curvature in quantum space.
In this framework, the eigenvalue equation (\ref{eg}) takes the form
\begin{equation}\label{eq2}
{\cal H}\psi=\left(\frac{2m E}{\hbar^2}+\alpha\right)\psi=\varepsilon_n\psi,
\end{equation}
where the Hamiltonian ${\cal H}$ is given by

\begin{equation}\label{H}
\mathcal{H} = -\big(1+\alpha x^2\big)^2 \frac{\rd^2}{\rd x^2} - 4\alpha x\big(1+\alpha x^2\big) \frac{\rd}{\rd x} - 2\alpha^2 x^2 + a x^2 + \gamma x + \frac{\beta_{\textrm{rep}}}{x^2}.
\end{equation}
To analyze the spectrum of this system, we adopt a factorizable approach inspired by SUSYQM, generalized to non-adjoint operators. We then introduce a pair of operators:
\begin{equation}
B = -F(x) \frac{\rd}{\rd x} + W(x) - \Omega(x), \quad A = F(x) \frac{\rd}{\rd x} + W(x) + \Omega(x),
\end{equation}
where the functions involved in the factorization are chosen as follows:
\begin{equation}
F(x) = 1 + \alpha x^2, \quad W(x) = b + c x + \frac{g}{x}, \quad \Omega(x) = 2\alpha x. \end{equation}
To highlight the supersymmetric structure of the deformed system, we consider here the factorization of two partner Hamiltonians,
\begin{equation}
{\cal H}_- \equiv {\cal H} = BA \quad \textrm{and} \quad
{\cal H}_+ = AB,
\end{equation}
of the form:
\begin{equation}
\mathcal{H}_{\mp} = -\left[F(x) \frac{\rd}{\rd x}\right]^2 - 2 F(x) \Omega(x) \frac{\rd}{\rd x} + V_{\mp}(x),
\end{equation}
with the corresponding potentials:
\begin{equation}
V_{\mp}(x) = c(c\mp\alpha)x^2 - 6\alpha^2x^2 + \frac{g(g\pm1)}{x^2} + \frac{2gb}{x} + 2bcx + b^2 + 2gc - \alpha g \mp c - 2\alpha.
\end{equation}
This construction links the system dynamics to a factorizable structure, making an analytical treatment of the spectrum possible.
The two partner Hamiltonians share a similar form, differing only by a shift in their parameters, which constitutes the essence of
supersymmetric quantum mechanics.

A crucial shape invariance condition imposes $b = 0$, thereby removing the $1/x$ term and implying $\gamma = 0$.
Shape invariance then manifests through the following relation between the partner potentials:
\begin{equation}
V_+(x; c; g) = V_-(x; c+\alpha; g-1) -4\alpha g + 4c + 4\alpha.
\end{equation}
This property ensures the preservation of the potential structure under supersymmetric transformation,
allowing for the recursive construction of the energy spectrum. The energy eigenvalues are then obtained as:
\begin{equation}
\varepsilon_n^- = 4\big(-n\alpha g + nc + n^2\alpha\big).
\end{equation}
The energy of the Hamiltonian ${\cal H}$ is given by $\varepsilon_n = \varepsilon_n^- + \varepsilon_0$,
with $\varepsilon_0 = -g(\alpha + 2c) + 2\alpha + c$, we obtain the explicit expression for the total energy of the state labeled by $n$:
\begin{equation}
E_n = \frac{\hbar^2}{2m}\left[(4n^2 + 1)\alpha + (4n + 1)(-\alpha g + c) - 2cg\right],
\end{equation}
where the parameters are defined as:
\begin{equation}
c = \frac{\alpha + \sqrt{\alpha^2 + 4(a + 4\alpha^2)}}{2}, \quad
g = \frac{-1 - \sqrt{1 + 4\beta_{\textrm{rep}}}}{2}, \quad
a = \frac{m^2\omega^2}{\hbar^2}.
\end{equation}

This approach emphasizes the key role of shape invariance: it not only ensures the internal consistency of the supersymmetric framework,
but also enables the exact determination of the energy spectrum, demonstrating the effectiveness of the SUSY factorization method
in the study of deformed quantum systems.

This solution describes a quantum oscillator influenced by three distinct physical contributions, whose interplay captures the effects of the deformed background:
\begin{itemize}
\item the effective curvature, introduced through the deformation, modifies the spatial metric and directly affects the dynamics of the system;
\item the harmonic term, derived from the potential $a x^2$, is responsible for the equally spaced energy levels in the flat limit;
\item the centrifugal term, governed by the parameter $\beta_{\textrm{rep}}$, introduces a repulsive contribution near the origin, preventing the particle from localizing too closely.
\end{itemize}

The resulting energy spectrum reveals that curvature induces nontrivial corrections depending on the quantum number $n$, in particular a quadratic term in $n^2$ that is absent in the standard case. These corrections become increasingly significant for highly excited states, illustrating the sensitivity of the system to the underlying geometric deformation and the breakdown of the uniform level spacing typical of the harmonic oscillator.

In the limit $\alpha \rightarrow 0$, the system reduces to a standard harmonic oscillator with an additional inverse-square potential \cite{Dong2007}, whose energy spectrum reads:
\begin{equation}
E_n=\hbar\omega\bigg(2n+1+\frac{1}{2}\sqrt{1+4\beta_{\textrm{rep}}}\bigg).
\end{equation}
The non-Hermitian SUSY factorization method has been adopted in the present study for three fundamental reasons: (i)~it yields an exact solution of the energy spectrum for any value of the deformation parameter $\alpha$, whereas stationary perturbation theory converges only in the limit $|\alpha| \ll 1$~\cite{Cooper1995}; (ii)~it unveils the hidden $\mathfrak{su}(1,1)$ symmetry of the deformed system, which remains completely inaccessible to perturbative methods~\cite{Dossa2020}; (iii)~it enables a unified treatment of both linear and quadratic deformations within a single formalism, while the perturbative approach would require independent developments for each deformation type, potentially with different radii of convergence. This analytical robustness proves indispensable for the thermodynamic analysis, which demands a precise knowledge of the energy levels~$E_n$ over an extended range of quantum numbers $n$.

The following section is devoted to the analysis of thermodynamic quantities associated with the quantum system under investigation. The objective is to elucidate how macroscopic properties emerge from the energy spectrum modified by geometric deformation, and to quantify the impact of this deformation on the  thermal behavior of the system at equilibrium.

\section{Determination of thermodynamic quantities}
\label{sec-5}

The thermodynamic study of a quantum system necessarily involves the analysis of its partition function, denoted $Z(\beta)$. This plays a central role: it contains all the statistical information of the system at thermal equilibrium and provides access to the fundamental thermodynamic quantities. For a discrete spectrum $\{E_n\}$, it is written:
\begin{equation}
Z(\beta) = \sum_{n=0}^{N_{\rm{max}}} \re^{-\beta E_n},\quad E_n={\cal A} n^2 + {\cal B} n + {\cal C}, \quad \beta = \frac{1}{k_{\rm B} T},
\end{equation}
where $k_{\rm B}$ denotes the Boltzmann constant, and $E_n$ is the energy levels of the system.

However, when the analytical expressions for $E_n$ become complex, the exact sum becomes difficult to evaluate. An efficient approach is to use the Euler--Maclaurin formula, which approximates a sum by an integral corrected by boundary terms:
\begin{equation}
\sum_{n=0}^{N} f(n) \approx \frac{1}{2}\big[f(0) + f(N+1)\big] + \int_0^{N+1} f(x) \,\rd x.
\end{equation}
This approximation offers an analytical compromise between accuracy and simplicity. It allows one to obtain a semi-analytic form of the partition function $Z(\beta)$, from which the following thermodynamic quantities can be calculated: the internal energy $U$, the Helmholtz free energy $F$, the entropy $S$, and the heat capacity $C$. We present this analysis below for two special cases of deformations.

\subsection*{a) Case of linear deformation: $\mu(x) = \alpha x$}

In this case, the influence of the linear deformation is manifested by an energy spectrum of the quadratic form:
\begin{equation}
E_n = A_1 n^2 + B_1 n + C_1,
\end{equation}
with the coefficients given by:
\begin{eqnarray}
A_1 = -\frac{\hbar^2 \alpha^2}{2m}, \quad
B_1 = -\frac{\hbar^2 \alpha^2}{2m} - \frac{\hbar \mathcal{E} \alpha}{m \omega} + \hbar \omega, \cr
C_1 = -\frac{\hbar \mathcal{E} \alpha}{2m \omega} + \frac{\hbar^2 \alpha^2}{4m} + \frac{\hbar \omega}{2} - \frac{\mathcal{E}^2}{2m \omega^2}.
\end{eqnarray}
Due to the quadratic form of the spectrum, the partition function can be evaluated analytically using the Euler--Maclaurin approximation. We obtain:
\begin{equation}
Z(\beta)=Z_1(\beta)+Z_2(\beta),
\end{equation}
where
\begin{equation}
Z_1(\beta) = \frac{1}{2} \left(\re^{-\beta C_1} + \re^{-\beta E_{{N}}}\right),
\end{equation}
\begin{equation}
Z_2(\beta) = \frac{1}{2} \sqrt{\frac{\piup}{-\beta A_1}} \re^{-\beta \left( C_1 - \frac{B_1^{2}}{4A_1} \right)} \left[ {\rm erfi}(X_1) - {\rm erfi}(X_2) \right],
\end{equation}
with
\begin{equation}
X_1 = \sqrt{-\beta A_1} \, u_1, \quad X_2 = \sqrt{-\beta A_1} \, u_2,
\end{equation}

\begin{equation}
u_1 = N+1 + \frac{B_1}{2A_1}, \quad u_2 = \frac{B_1}{2A_1}, \quad E_N = A_1(N+1)^2 + B_1(N+1) + C_1.
\end{equation}
${\rm erfi}$ is the imaginary error function given by
\begin{equation}
 {\rm erfi}(x)=\frac{2}{\sqrt{\piup}}\int_0^x \re^{t^2}\rd t.
\end{equation}

\subsection*{b) Case of a quadratic deformation: $\mu(x) = \alpha x^2$}

In this second case, the spectrum also exhibits a quadratic dependence:
\begin{equation}
E_n = A_2 n^2 + B_2 n + C_2,
\end{equation}
with
\begin{eqnarray}
A_2 = \frac{2\hbar^2 \alpha}{m}, \quad
B_2 = \frac{\hbar^2}{2m}(-\alpha g + c), \cr
C_2 = \frac{\hbar^2}{2m} \left[ \alpha (1 - g) + c(1 - 2g) \right].
\end{eqnarray}
The partition function obtained is:
\begin{equation}
Z(\beta)=Z_1(\beta)+Z_2(\beta),
\end{equation}
where
\begin{equation}
Z_1(\beta) = \frac{1}{2} \left[ \re^{-\beta C_2} + \re^{-\beta E_{{N}}}\right],
\end{equation}
\begin{equation}
Z_2(\beta) = \frac{1}{2} \sqrt{\frac{\piup}{\beta A_2}} \re^{-\beta \left( C_2 - \frac{B_2^{2}}{4A_2} \right)} \left[ \textrm{erf}(Y_1) - \textrm{erf}(Y_2) \right],
\end{equation}
with
\begin{equation}
Y_1 = \sqrt{\beta A_2} \, v_1, \quad Y_2 = \sqrt{\beta A_2} \, v_2,
\end{equation}
\begin{equation}
v_1 = N+1 + \frac{B_2}{2A_2}, \quad v_2 = \frac{B_2}{2A_2}, \quad E_N = A_2(N+1)^2 + B_2(N+1) + C_2.
\end{equation}
$\textrm{erf}$ is the real error function given by
\begin{equation}
 \textrm{erf}(x)=\frac{2}{\sqrt{\piup}}\int_0^x \re^{-t^2}\rd t.
\end{equation}

\subsection*{c) Derivation of thermodynamic quantities}

In both deformation scenarios discussed above, the thermodynamic observables are obtained directly from the canonical partition function $Z(\beta)$. The main quantities of interest are given by the standard relations:

\textbf{Internal energy:}
\begin{equation}
U(\beta) = -\frac{\partial}{\partial \beta} \ln Z(\beta),
\end{equation}
or equivalently,
\begin{equation}
U(\beta) = -\frac{1}{Z(\beta)} \left( \frac{\rd Z_1}{\rd\beta} + \frac{\rd Z_2}{\rd\beta} \right),
\end{equation}
where $Z_1$ and $Z_2$ denote the respective contributions to the partition function.

\textbf{Helmholtz free energy:}
\begin{equation}
F(\beta) = -\frac{1}{\beta} \ln Z(\beta).
\end{equation}

\textbf{Entropy:}
\begin{equation}
S(\beta) = k_{\rm B} \ln Z(\beta) + k_{\rm B} \beta U(\beta).
\end{equation}

\textbf{Heat capacity:}
\begin{equation}
C(\beta) = -k_{\rm B} \beta^2 \frac{\rd U}{\rd \beta}.
\end{equation}

These expressions provide a complete thermodynamic characterization of the system and allow us to analyze how geometric deformations modify its thermal behavior. In particular, they make it possible to identify signatures of the effective spatial geometry at the level of quantum thermodynamics. The following section is devoted to the numerical evaluation of these quantities and to the physical interpretation of the resulting thermal profiles.

\section{Numerical results}
\label{sec-6}

In this section, we present the numerical results obtained for the two deformation cases studied: linear $\mu(x) = \alpha x$ and quadratic $\mu(x) = \alpha x^2$. We first analyze the evolution of energy levels as a function of the deformation parameter $\alpha$. We then explore the combined effect of temperature $T$ (or the thermal inverse $\beta = 1/k_{\rm B} T$) and deformation on thermodynamic properties such as internal energy $U$, free energy $F$, entropy $S$, and heat capacity $C$. In order to simplify numerical simulations and ensure a consistent comparison between linear and quadratic deformation regimes, the fundamental constants are fixed according to the following system of natural units:
\begin{equation}
 \hbar=1,\quad m=1,\quad \omega=1,\quad {\cal E}=0, \quad \beta_{\rm rep}=0.
\end{equation}

\begin{figure}[ht]
	\centering
	\includegraphics[scale=0.42]{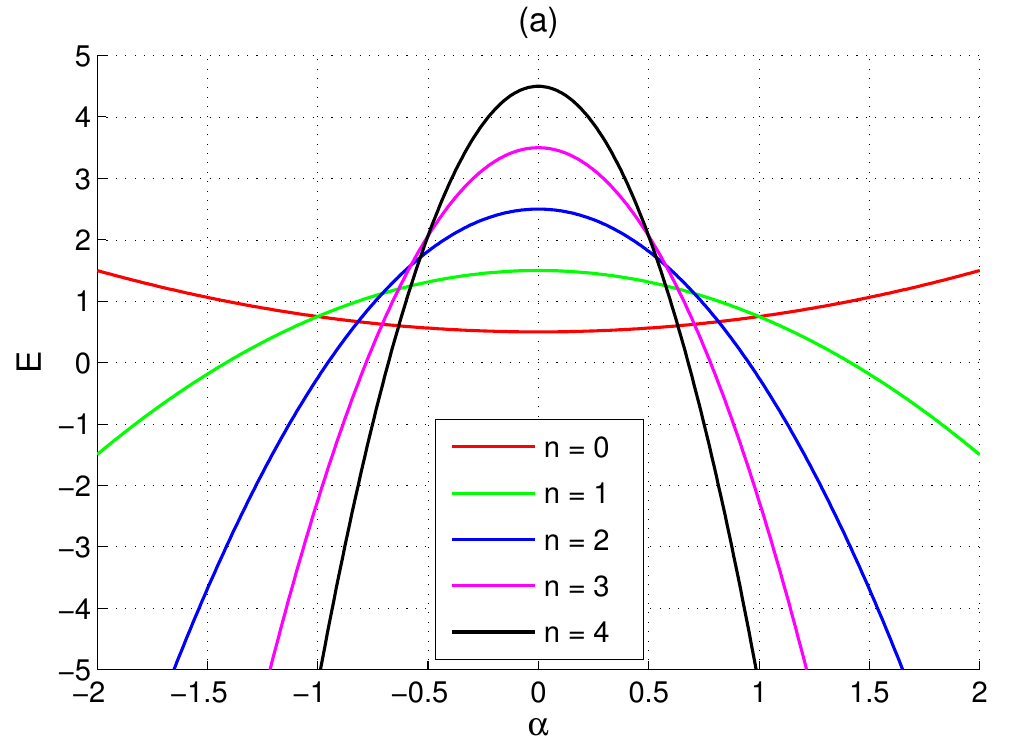}
	\includegraphics[scale=0.48]{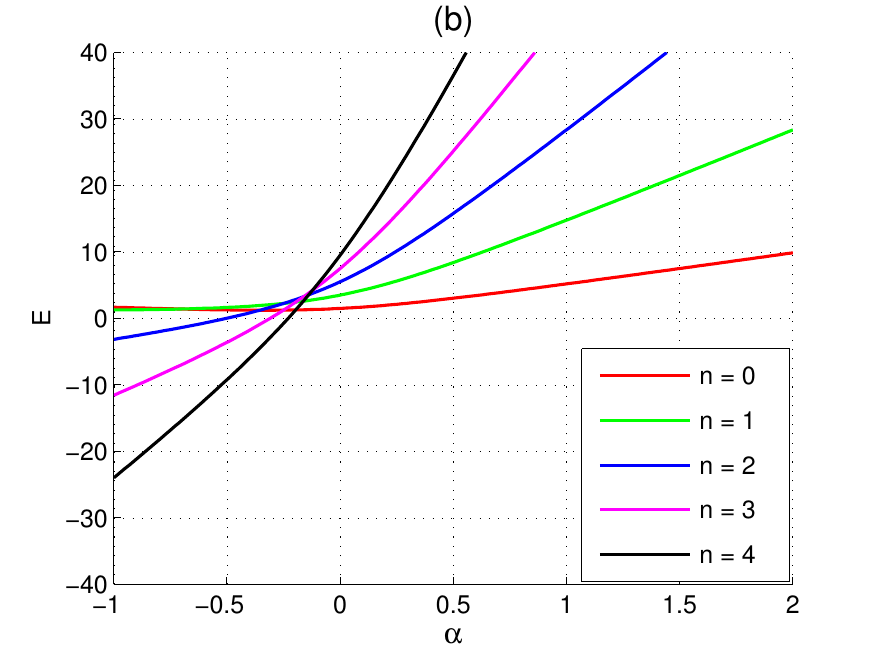}
	\caption{(Colour online) Evolution of energy levels as a function of the deformation parameter $\alpha$: (a)~linear deformation; (b) quadratic deformation.}
	\label{fig:fig2a}
\end{figure}

Figure \ref{fig:fig2a} illustrates the evolution of energy levels as a function of the deformation parameter $\alpha$. In the linear case figure (a), we observe that for low values of $|\alpha|$, all excited energy levels remain positive, while the fundamental level is strictly positive regardless of the value of $\alpha$. This suggests a stability of the spectrum for small deformations. On the other hand, in the quadratic case figure (b), all energy levels remain positive as long as $\alpha > 0$, which confirms that this type of deformation induces a more regular and well-defined growth of the energy spectrum. This spectral analysis is an essential prerequisite for the study of thermodynamic properties, since the equilibrium probability distributions are directly linked to the structure of the energy levels.

\begin{figure}[ht]
\centering
\includegraphics[width=0.46\textwidth]{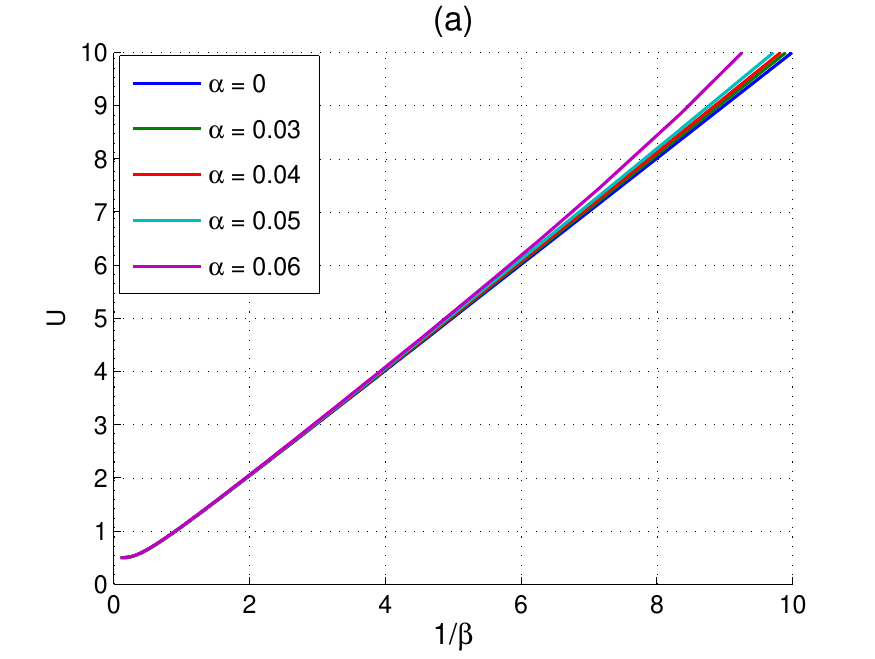}
\includegraphics[width=0.46\textwidth]{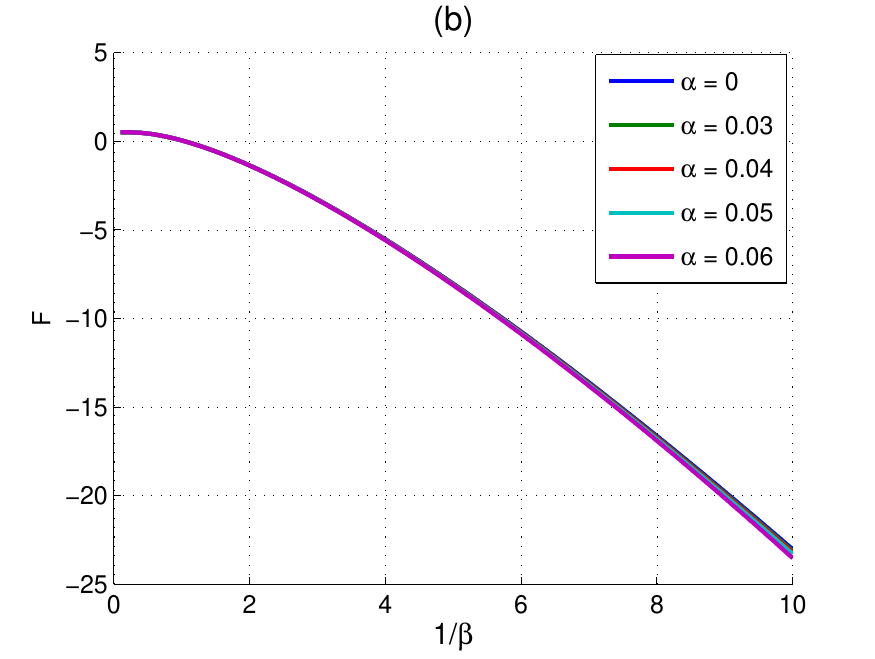}
\includegraphics[width=0.46\textwidth]{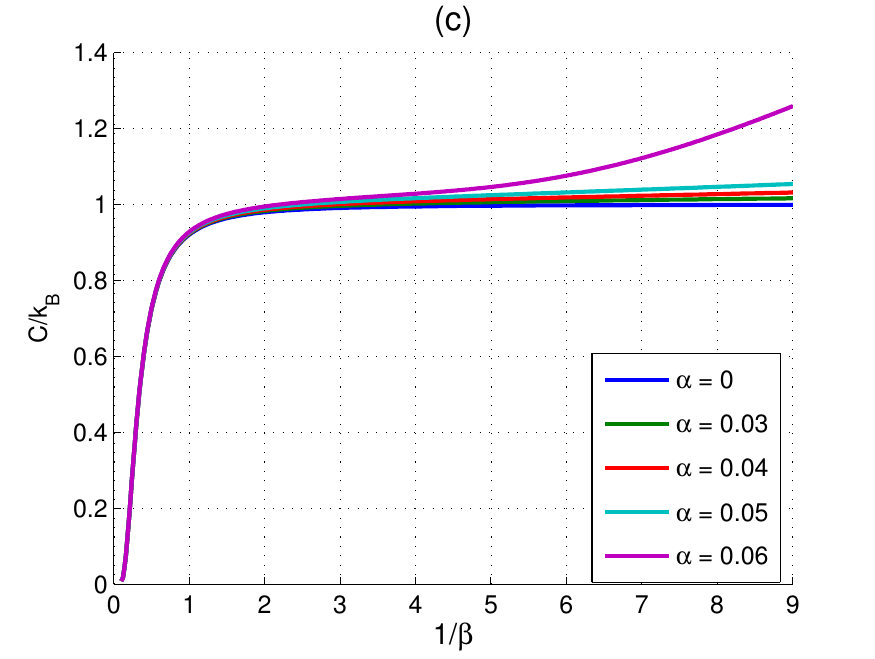}
\includegraphics[width=0.46\textwidth]{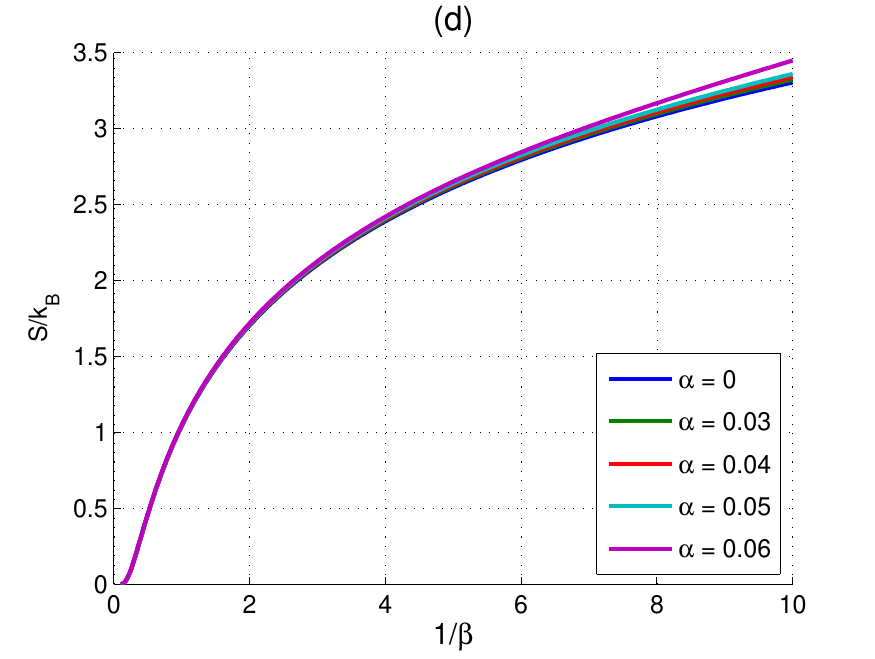}
\caption{(Colour online) Thermodynamic properties as a function of temperature for different values of~$\alpha$ (case of linear deformation).}
\label{fig:fig2b}
\end{figure}

Figure \ref{fig:fig2b} illustrates the thermodynamic properties of the system subjected to linear deformation.
The internal energy ($U$) exhibits a monotonous growth with temperature. This increase becomes more pronounced as the deformation parameter $\alpha$ increases, reflecting an increase in the number of accessible energy states at high temperatures.
Simultaneously, the free energy ($F$) decreases with increasing temperature, which is characteristic of a thermally excited system. The effect of the deformation is manifested by a change in the slope of this decrease, highlighting the influence of $\alpha$ on the thermodynamic potential of the system.

The modification of the density of states $\rho(E) = |\rd n/\rd E|$ induced by geometric deformation accounts for the singularities observed in thermodynamic properties. For the standard harmonic oscillator, $\rho_0(E) = 1/\hbar\omega$ is constant.

In the case of a linear deformation $\mu(x)=\alpha x$, the coefficient $A_1 = -\hbar^2\alpha^2/(2m) < 0$ leads to a concave parabolic spectrum with a maximum energy $E_{\max}$. The associated density of states diverges as:
\[
\rho_\alpha(E) \propto \frac{1}{\sqrt{E_{\max} - E}}, \quad (E < E_{\max}),
\]
which is responsible for the peak in heat capacity observed for $\alpha = 0.06$ [figure~\ref{fig:fig2b}(c)]. This divergence reflects an accumulation of states near the maximum accessible energy, a phenomenon characteristic of systems with an upper-bounded spectrum.

\begin{figure}[ht]
\centering
\includegraphics[width=0.48\textwidth]{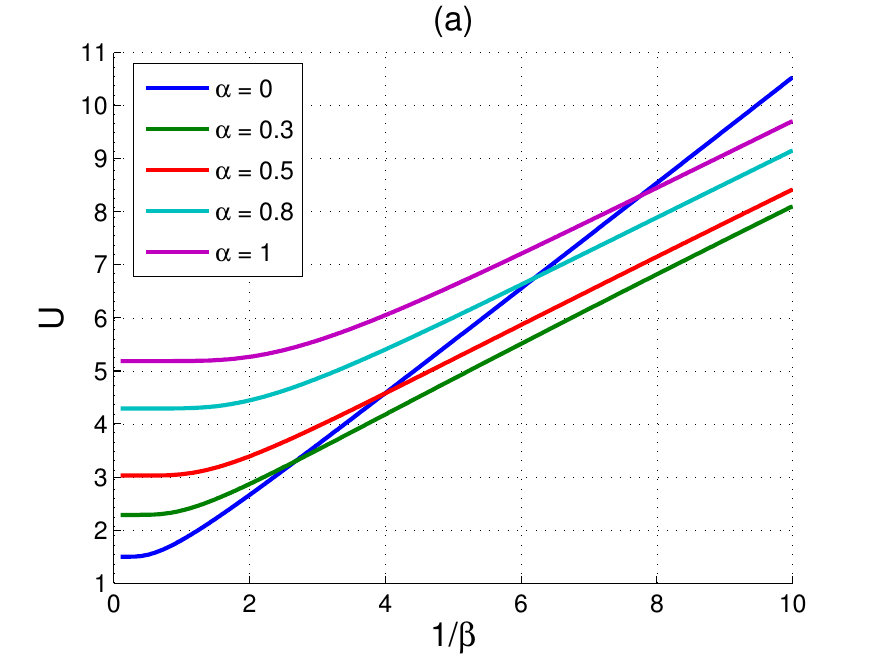}
\includegraphics[width=0.48\textwidth]{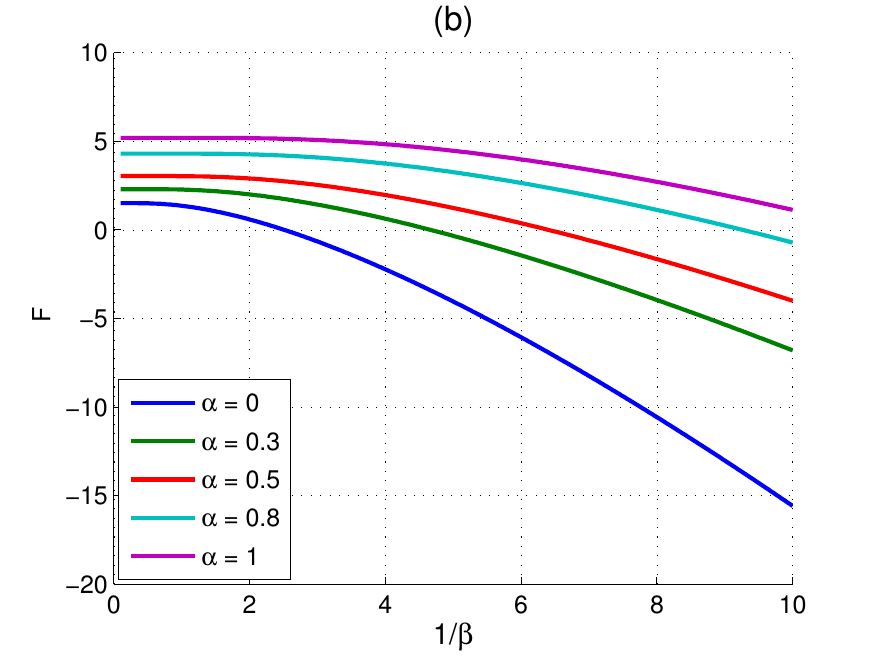}
\includegraphics[width=0.48\textwidth]{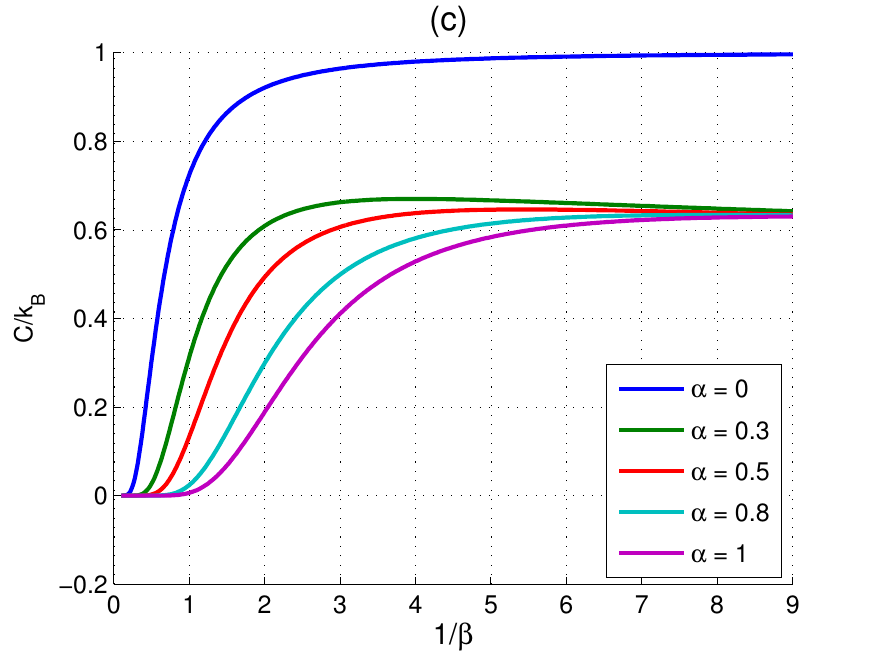}
\includegraphics[width=0.48\textwidth]{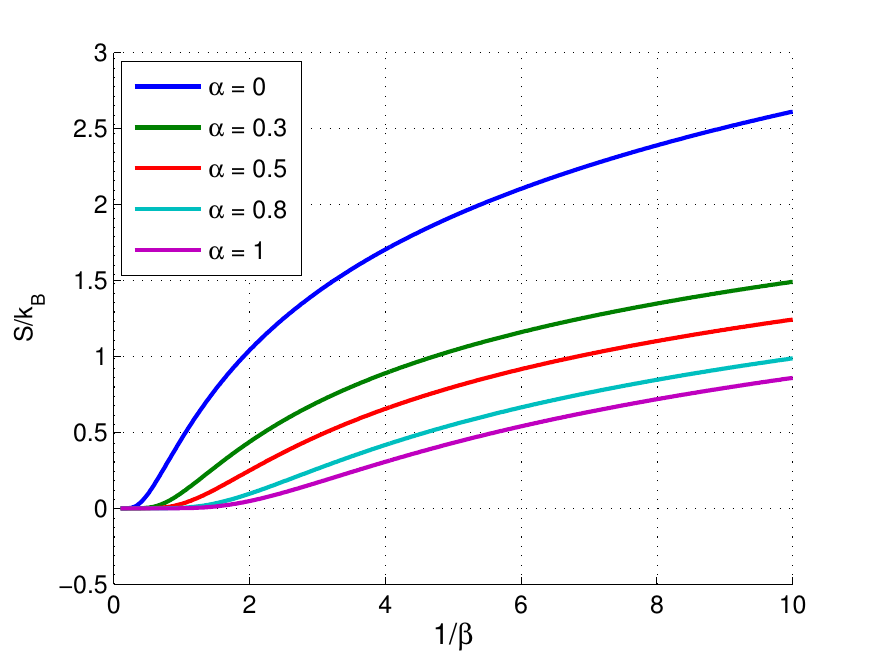}
\caption{(Colour online) Thermodynamic properties as a function of temperature for different values of~$\alpha$ (case of quadratic deformation).}
\label{fig:fig2c}
\end{figure}

For the quadratic deformation $\mu(x)=\alpha x^2$, the coefficient $A_2 = 2\hbar^2\alpha/m > 0$ ($\alpha>0$) induces a spectrum that grows quadratically, $E_n \propto n^2$, for large $n$, which modifies the density of states to $\rho_\alpha(E) \propto 1/\sqrt{E}$. This decay with energy limits the number of thermally accessible states at high temperature, explaining the observed saturation of the heat capacity toward $C/k_{\rm B} \approx 0.6$ [figure~\ref{fig:fig2c}(c)], in contrast with the Dulong--Petit law ($C/k_{\rm B} \to 1$) of the standard oscillator. This behavior could model geometric confinement effects in curved carbon nanotubes~\cite{Serra1997} or spherical quantum dots~\cite{Gangopadhyaya2018,Dossa2021}.

Finally, entropy ($S$) increases with temperature, reflecting an increase in disorder. This increase becomes more rapid when $\alpha$ is higher, suggesting that deformation promotes accelerated access to an increasing number of accessible states.

Figure \ref{fig:fig2c} presents the thermodynamic properties of the system subjected to quadratic deformation. Although trends similar to those observed for linear deformation are identifiable, notable differences emerge, particularly in the intensity of the effects induced by the deformation.

Thus, the internal energy ($U$) maintains its increasing behavior with temperature. However, the influence of the deformation parameter $\alpha$ is more pronounced in this case, particularly at low temperatures, where significant differences between the curves are observed. This indicates an increased sensitivity of the system to quadratic deformation in weak thermal regimes.

The free energy ($F$), for its part, continues to decrease with increasing temperature. However, the effect of quadratic deformation is more pronounced than in the linear case, more significantly modifying the amount of energy available to perform thermodynamic work.

The heat capacity ($C$) exhibits a particularly interesting behavior. While at low temperatures it is strongly affected by deformation, at high temperatures it becomes almost insensitive to the variation of~$\alpha$. It then tends towards a limiting value, approximately $1$ for $\alpha = 0$ and $0.6$ for $\alpha \neq 0$, suggesting a saturation of the number of accessible states. This saturation can be interpreted as the consequence of a modified density of states, with the quadratic deformation limiting the energy access to the excited states of the system.

Finally, the entropy ($S$) continues to increase with temperature, but its growth is significantly slowed in the presence of a quadratic deformation. This attenuation reflects a more gradual access to disordered states, consistent with a narrower energy spectrum for $\alpha > 0$.

In conclusion, these numerical results highlight the structural impact of the deformation shape on the thermodynamic behavior of the system. While linear deformation allows a smoother transition between states, quadratic deformation induces a stiffening of the energy spectrum, resulting in more marked and less gradual thermodynamic effects.

\section{Conclusion}
\label{sec-7}

In this work, we have developed a comprehensive theoretical framework for quantum systems with position-dependent geometric deformations, based on the generalized extended momentum operator that ensures strict self-adjointness of the deformed momentum. Unlike the approaches relying solely on boundary-condition hermiticity~\cite{Bagchi2005}, our formalism ensures intrinsic self-adjointness through the unique GEMO structure, thereby preserving a probability conservation in curved effective geometries.

We have systematically analyzed both linear ($\mu(x)=\alpha x$) and quadratic ($\mu(x)=\alpha x^2$) deformations within a unified non-Hermitian supersymmetric framework. Although the factorization employs non-adjoint ladder operators, the resulting Hamiltonian remains rigorously Hermitian, enabling exact spectral solutions while revealing hidden $\mathfrak{su}(1,1)$ symmetry structures inaccessible to perturbative methods.
Crucially, our analysis extends beyond spectral properties addressing a gap left by previous works~\cite{Bagchi2005} by providing the first thermodynamic characterization of GEMO-based systems. Through the Euler--Maclaurin approximation of the partition function, we have established a direct link between geometric deformation and thermal signatures, identifying the modified density of states
$\rho(E)$ as the microscopic origin of anomalous heat capacity behaviors: a divergence $\rho(E)\propto 1/\sqrt{E_{\rm max}-E}$ for linear deformation and an algebraic decay $\rho(E)\propto 1/\sqrt{E}$ for quadratic deformation.

Our numerical results demonstrate that geometric deformation acts as an effective control parameter for quantum thermodynamics. For $\alpha>0$, spectral stability is preserved while thermodynamic responses are profoundly reshaped: linear deformation induces pronounced peaks in heat capacity due to state accumulation near $E_{\rm max}$, whereas quadratic deformation leads to saturation of $C/k_{\rm B}\to 0.6$ at high temperature departing from the Dulong--Petit limit ($C/k_{\rm B}\to 1$) of the standard oscillator. These signatures may prove experimentally relevant for curved nanostructures such as carbon nanotubes~\cite{Serra1997} or spherical quantum dots~\cite{Gangopadhyaya2018,Dossa2021}, where effective geometry directly influences the thermal transport.

The present study considered a static electric field $E$ to preserve exact solvability within the SUSYQM framework. An extension to time-dependent fields $E(t)=E_0\cos(\Omega t)$ \cite{Grifoni1998}
would be physically relevant for modelling the driven-dissipative dynamics in mesoscopic
systems.
However, such generalization breaks the time-translation invariance, rendering standard SUSY factorization inapplicable. Future work could address this challenge through Floquet theory for periodic driving or master-equation approaches incorporating environmental coupling both representing promising avenues beyond the exact-solvability regime explored here.

In summary, this work establishes geometric deformation via GEMO as a powerful tool for engineering quantum thermodynamic responses. The synergy between strict self-adjointness, exact spectral solutions, and analytical thermodynamics provides a robust foundation for exploring more complex scenarios, including many-body interactions, topological deformations, and non-equilibrium quantum thermal machines.



\newpage

\ukrainianpart

\title{Спектральні та термодинамічні властивості суперсиметричних квантових систем із самоспряженим деформованим імпульсом}
\author{Й. А. Оке\refaddr{label1}, Ф. А. Досса\refaddr{label2,label3}}
\addresses{
	\addr{label1} Школа наук, технологій, інженерії та математики (ED-STIM), Національний університет наук, технологій, інженерії та математики (UNSTIM), Бенін
	\addr{label2} Лабораторія фізики та прикладних застосувань університетського центру Натітінгу,   Бенін
	\addr{label3} Кафедра фізики, факультет науки і техніки (FAST/UNSTIM), Бенiн
}

\makeukrtitle

\begin{abstract}
	\tolerance=3000%
	На основі строгого підходу до опису квантових систем з геометричними деформаціями, побудовано самоспряжений деформований оператор імпульсу за допомогою формалізму узагальненого розширеного оператора імпульсу. На відміну від попередніх підходів, що ґрунтуются на ермітовості граничних умов, наш метод забезпечує внутрішню самоспряженість як для лінійних ($\mu(x)=\alpha x$), так і для квадратичних ($\mu(x)=\alpha x^{2}$) деформацій в рамках уніфікованої неермітової схеми суперсиметричної факторизації. Це забезпечує точні аналітичні спектри, водночас виявляючи приховані структури симетрії $\mathfrak{su}(1,1)$. Найважливіше те, що ми досліджуємо термодинамiчні характеристики таких систем шляхом аналітичної оцінки функції розподілу за допомогою наближення Ейлера-Маклорена. Геометрична деформація фундаментально змінює густину станів $\rho(E)$, створюючи чіткі теплові сигнатури: розбіжний пік теплоємності для лінійної деформації через накопичення станів поблизу максимальної енергії та насичення $C/k_{\mathrm{B}}\to 0.6$ (нижче межі Дюлонга--Пті) для квадратичної деформації. Ці результати характеризують геометричну деформацію як параметричне налаштування для можливого квантового термодинамічного відгуку у деформованих наноструктурах.
	\keywords суперсиметрія, самоспряжений імпульс, термодинамічні властивості
	
\end{abstract}

\lastpage
\end{document}